\input{aipcheck}

\documentclass[
    ,final            
  ]
  {aipproc}

\layoutstyle{8x11double}
\usepackage{bm}
\usepackage{amsmath}
\usepackage[abs]{overpic}

\begin{document}

\title{Simulation of pattern dynamics of cohesive granular particles
under a plane shear}


\author{Satoshi Takada}{}

\author{Hisao Hayakawa}{
address={Yukawa Institute for Theoretical Physics, Kyoto University,
Kitashirakawa Oiwakecho, Sakyo-ku, Kyoto 606-8502 Japan}
}


\begin{abstract}
We have performed three-dimensional molecular dynamics simulation of
cohesive granular particles under a plane shear.
From the simulation, we found the existence of three distinct phases in steady states: (I) a uniform
shear phase, (II) a coexistent phase of a shear band and a gas region and (III)
a crystal phase. We also found that the critical line between
(II) and
(III) is approximately represented by $\zeta \propto \exp(\beta
\dot\gamma L_y)$,
where $\zeta$, $\beta$, $\dot\gamma$, $L_y$ are the dissipation rate,
an unimportant constant, the shear rate and the system size of the
velocity gradient direction, respectively.
\end{abstract}

\maketitle


\section{Introduction}
\quad The interactions among macroscopic granular particles, such as sands,
are characterized by a repulsive and a dissipative ones.
The energy dissipation through inelastic collisions causes destabilization
of a uniform state of the system.
It is known that there appear clusters in such systems \cite{Goldhirsch,
Goldhirsch2},
which may be understood by the hydrodynamic equations \cite{McNamara, McNamara2, Brilliantov, Brilliantov2}.
When a shear is applied to the granular system,
high-density region, called "shear band", appears
\cite{Saitoh2007}.
There exist many papers to estimate the transport coefficients by using
kinetic theory
\cite{Jenkins1983,Lutsko2005,Garzo1999,Lun1984,Sela1996,
Montanero1999} and to analyze the pattern dynamics by using continuum
mechanics
\cite{Saitoh2011,Alam2012,Alam1998,Alam1997,Savage1992,Garzo2006,Schmid1994,Gayen2006,Nott1999}.

On the other hand, fine powders of submicron order,
such as tonner particles or interstellar dusts, have attractive
forces such as an electrostatic force \cite{Castellanos}.
The existence of the attractive force causes some new features due to
the competition of the gas-liquid phase transition \cite{Hansen, Smit,
Watanabe}
and the dissipative structure \cite{Saitoh2007}.
For instance, nucleation process near equilibrium is well understood
\cite{Yasuoka},
but that under a shear is not well understood.
In this paper, we try to characterize the non-equilibrium pattern
formation of fine powders under a plane shear
based on the three dimensional molecular dynamics simulation.

\section{Model and Setup}
\quad The system we consider consists of monodisperse $N(=10,000)$ spheres, whose radius
is $\sigma$ and mass is $m$.
The system size is $L_x \times L_y \times L_z$ $(L \equiv L_x = L_y)$ and
we choose $x$-axis as the shear direction and $y$-axis as the velocity
gradient
direction.
We assume that the interaction between particles is described 
by the truncated Lennard-Jones potential
\begin{equation}
U^{\rm LJ}(r_{ij})=
\begin{cases}
4\varepsilon\left( \left( \frac{\sigma}{r_{ij}} \right)^{12} - \left(
\frac{\sigma}{r_{ij}} \right)^{6} \right) & (r\leq r_{\rm c})\\
0 & (\rm else)
\end{cases},
\end{equation}
with the well depth $\varepsilon$,
where $r_{ij}=|\bm{r}_{ij}|$ is the distance between $i$-th and $j$-th
particles
and $r_{\rm c}$ is cut-off length (in this study we use $r_{\rm
c}=3\sigma$).
For the dissipative force, we use
\begin{equation}
\bm{F}^{\rm vis}(\bm{r}_{ij},\bm{v}_{ij})=-\zeta
\Theta(\sigma-|\bm{r}_{ij}|)(\bm{v}_{ij} \cdot \hat{\bm{r}}_{ij})
\hat{\bm{r}}_{ij},
\end{equation}
with the dissipation rate $\zeta$, where $\bm{v}_{ij}$ is the relative
velocity vector of $i$-th and $j$-th particles,
$\Theta(r)$ is the step function which is $1$ for $r>0$ and $0$ for otherwise,
and $\hat{\bm{r}}$ is a unit vector proportional to $\bm{r}$.
We note that $\zeta$ is a dissipation parameter related to the coefficient
of restitution $e$, for example, $e=0.998$ for $\zeta=0.1$ and $e=0.983$ for $\zeta=1.0$
at the temperature $T=1.4\varepsilon$.
Thus, we are interested in weakly dissipative situations.
This weak dissipation is necessary to reach a steady state.

In general, the boundary effect is important for granular systems.
It is known that
the results strongly depend on the boundary condition
\cite{Karison1999,Richman1988, Dave1995,Campbell1997,Tan1997}.
In this paper, we adopt Lees-Edwards condition
\cite{LEbc, Evans} to suppress the boundary effect
with the aid of SLLOD algorithm \cite{Evans, SLLOD}:
\begin{eqnarray}
\frac{d\bm{r}_i}{dt} &=& \frac{\bm{p}_i}{m} + \dot\gamma y_i
\hat{\bm{e}}_x,\\
\frac{d\bm{p}_i}{dt} &=& \bm{F}_i - \dot\gamma p_{yi}\hat{\bm{e}}_x.
\end{eqnarray}
This model is known as that the system is relaxed to the uniform shear
state near equilibrium,
where $\bm{r}_i=(x_i, y_i, z_i), \bm{p}_i=(p_{xi},p_{yi},p_{zi})$ are
respectively
the position and the momentum of $i$-th $(1\leq i\leq N)$ particle,
$\bm{v}_i=\dot{\bm{r}}_i$,
$\dot\gamma$ is the shear rate
and $\hat{\bm{e}}_x$ is the unit vector in $x$ direction.
The force acting on $i$-th particle is given by
\begin{equation}
\bm{F}_i = -\sum_{j \neq i} \bm{\nabla}_i U^{\rm LJ}(r_{ij}) + \sum_{j \neq
i}\bm{F}^{\rm vis}(\bm{r}_{ij}, \bm{v}_{ij}).
\end{equation}
In this paper, we choose $T_0=1.4 \varepsilon$ as the initial
temperature, which is slightly higher than the critical temperature for the
equilibrium Lennard-Jones fluid \cite{Hansen, Smit}.
The system is equilibrated in the absence of the shear and the dissipation
for $t=200 (m\sigma^2/\varepsilon)^{1/2}$.
Then we apply the shear and the dissipation to the system.

\section{Results}
\begin{figure}
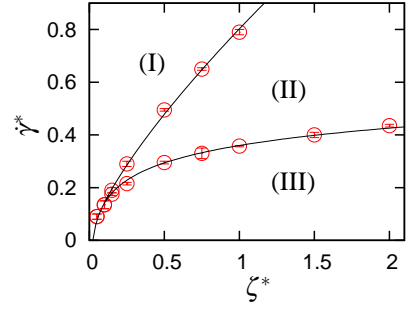

\begin{overpic}[height=.18\textheight]{phasediagram0308}
\put(90,-5){$\zeta^\ast$}
\put(3,55){\rotatebox{90}{${\dot\gamma}^\ast$}}
\put(50,80){(I)}
\put(100,70){(II)}
\put(100,35){(III)}
\end{overpic}
\caption{The phase boundary lines in the plane of the shear rate $\dot\gamma$ and the
dissipation rate $\zeta$ for the density $\bar\rho=0.308$.}
\label{fig:phase0308}
\end{figure}

\quad At first, we choose the average density $\bar\rho=0.308$,
where $\bar\rho$ is defined by $N\sigma^3/L^2L_z$.
Note that this density corresponds to the critical density for
Lennard-Jones fluid at
equilibrium \cite{Hansen, Smit, Watanabe}.
We fix the system size $L_x=L_y=52\sigma, L_z=12\sigma$ for this density.
We study how the steady state depends on parameters such as the density,
the shear rate and the dissipation rate.
\begin{figure}
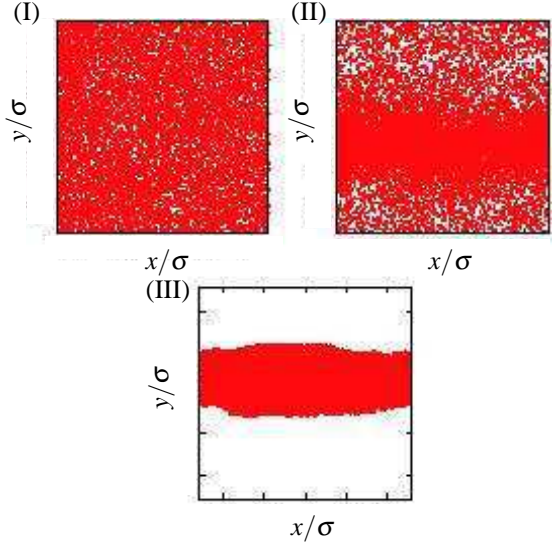

\begin{overpic}[height=.3\textheight]{steady_2}
\put(100,-5){$x/\sigma$}
\put(46,43){\rotatebox{90}{$y/\sigma$}}
\put(45,96){$x/\sigma$}
\put(-8,144){\rotatebox{90}{$y/\sigma$}}
\put(152,96){$x/\sigma$}
\put(99,144){\rotatebox{90}{$y/\sigma$}}
\put(-5,190){(I)}
\put(100,190){(II)}
\put(45,85){(III)}
\end{overpic}
\caption{Three typical patterns corresponding to three phases in steady states for $\bar\rho=0.308$: (I)
uniform shear phase (${\dot
\gamma}^\ast=0.88,\zeta^\ast=1.0$), (II) coexistent phase of shear band
and gas (${\dot\gamma}^\ast=0.66,\zeta^\ast=1.0$), and (III) crystal
phase (${\dot
\gamma}^\ast=0.355,\zeta^\ast=1.0$).}
\label{fig:pattern}
\end{figure}
We found that there exist three distinct steady phases.
The typical patterns for these phases are drawn in Fig.
\ref{fig:pattern},
where ${\dot\gamma}^\ast = \dot\gamma (m\sigma^2/\varepsilon)^{1/2}$ and
$\zeta^\ast = \zeta (m\sigma^2/\varepsilon)^{1/2}$.
The uniform shear phase (I) can be observed for a larger shear
rate,
and the crystal phase (III) is obtained for smaller shear rate.
We found that the coexistence phase (II) between a shear band and the gas region disappears
when the shear rate $\dot\gamma$ and the dissipative rate $\zeta$ become
small. The phase boundary lines for $\bar\rho=0.308$ are plotted in Fig.
\ref{fig:phase0308}.

Now let us reproduce the boundary lines between two phases.
For this purpose, we try to characterize the behavior
in terms of a simple physical argument based on the granular temperature $T_{\rm
g}\equiv(m/3N)\sum_{i=1}^N | \bm{v}_i - \bar{\bm{v}} |^2$
\cite{Goldhirsch2008,Goldhirsch2003},
where $\bar{\bm{v}}=\bar{\bm{v}}(y,t)$ is the velocity field of the $y$-axis.
We measure the time averaged granular temperature $\bar{T}_{\rm g}$, changing the shear rate $\dot\gamma$.
\begin{figure}
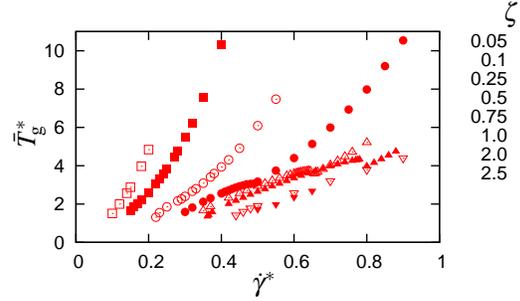

\begin{overpic}[height=.16\textheight]{Tg_gamma}
\put(90,-4){${\dot\gamma}^\ast$}
\put(0,50){\rotatebox{90}{$\bar{T}_{\rm g}^\ast$}}
\put(185,98){$\zeta$}
\end{overpic}
\caption{The dependence of the average granular temperaure $\bar{T}_{\rm g}$ on the shear rate
$\dot\gamma$ for several
dissipation rates $\zeta$ (for $\bar\rho=0.308$).}
\label{fig:Tg_gamma}
\end{figure}

The dependence of $\bar{T}_{\rm g}$
on the shear rate $\dot\gamma$ for several dissipation rates $\zeta$ is showed in Fig. \ref{fig:Tg_gamma},
where $\bar{T}_{\rm g}^\ast$ is defined by $\bar{T}_{\rm g}^\ast =
\bar{T}_{\rm g}/\varepsilon$.
In the case of the phase (I), the granular temperature satisfies $\bar
T_{\rm g} \propto \dot\gamma^2$,
which is consistent with the results of dimensional analysis
\cite{Goldhirsch2003, Haff1983}.
On the contrary, the granular temperature
$T_{\rm g}$ becomes zero as time goes on, i.e. $\bar{T}_{\rm g} =0$ in the phase (III).

\begin{figure}
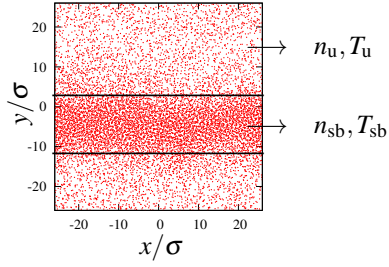

\begin{overpic}[height=.15\textheight]{steady2}
\put(65,-6){$x/\sigma$}
\put(15,43){\rotatebox{90}{$y/\sigma$}}
\put(32,30){\bf{------------------------}}
\put(32,52){\bf{------------------------}}
\put(105,40){${\bf{\longrightarrow}} \quad n_{\rm sb}, T_{\rm sb}$}
\put(105,70){${\bf{\longrightarrow}} \quad n_{\rm u}, T_{\rm u}$}
\end{overpic}
\caption{Typical configuration in pattern (II) ($\bar\rho=0.308,
{\dot\gamma}^\ast=0.66, \zeta^\ast=1.0$).
The system is separated into two regions as a function of the local
density $\rho(y)$ and the average density $\bar\rho$: (i) the shear band
region, and (ii) the gas region.}
\label{fig:steadyII}
\end{figure}

Here, we investigate the dependence of $\bar{T}_{\rm g}$ on $\dot\gamma$ in the phase (II).
A typical configuration of the particles is presented in Fig.
\ref{fig:steadyII}.
We can separate the system into two regions: (i) the shear band region, and
(ii) the gas region.

To this end, we consider the density profile $\rho(y)$, which is defined
as follows: we count the number of particles in the region $(j-1)\sigma
< y < j\sigma$, where $j$ is an integer, and calculate the local average
density $\rho_j$ in that region. If
$\rho_j > \bar\rho$, this region is regarded as a part of (i),
otherwise the region is (ii).
We introduce $n_{\rm sb}$ and $T_{\rm sb}$ as the number and the
granular temperature in
the region (i) respectively. $n_{\rm u}$ and $T_{\rm u}$
are the corresponding ones in the
region (ii).
The average granular temperature of the system $\bar{T}_{\rm g}$ is given by
\begin{eqnarray}
\bar{T}_{\rm g} = \frac{N-n_{\rm u}}{N}T_{\rm sb} + \frac{n_{\rm u}}{N}T_{\rm
u} \label{eq:Tg}.
\end{eqnarray}
\begin{figure}
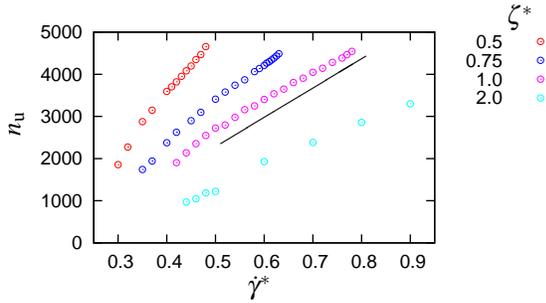

\begin{overpic}[height=.16\textheight]{number}
\put(80,52){\line(5,3){55}}
\put(90,-5){${\dot\gamma}^\ast$}
\put(0,55){\rotatebox{90}{$n_{\rm u}$}}
\put(188,97){$\zeta^\ast$}
\end{overpic}
\caption{The number of particles in the region (ii) $n_{\rm u}$ vs. the
shear rate $\dot\gamma$.}
\label{fig:number}
\end{figure}
For $\bar\rho=0.308$, the relationship between the shear rate
$\dot\gamma$ and the number of particles in region (ii) $n_{\rm u}$ is
plotted
in Fig. \ref{fig:number}.
From Fig. \ref{fig:number}, we approximately found $n_{\rm u} \propto
\dot\gamma$.
\begin{figure}
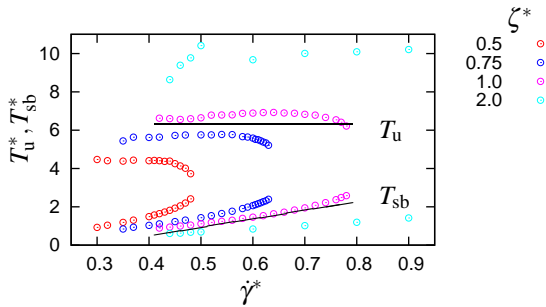

\begin{overpic}[height=.16\textheight]{temperature}
\put(55,60){\line(1,0){75}}
\put(55,18){\line(6,1){75}}
\put(87,-6){${\dot\gamma}^\ast$}
\put(0,45){\rotatebox{90}{$T_{\rm u}^\ast,T_{\rm sb}^\ast$}}
\put(140,55){$T_{\rm u}$}
\put(140,30){$T_{\rm sb}$}
\put(188,97){$\zeta^\ast$}
\end{overpic}
\caption{The shear rate $\dot\gamma$ vs. the granular temperature
$T_{\rm sb}$ in the region (i)
and $T_{\rm u}$ in the region (ii).}
\label{fig:temp}
\end{figure}
The relationships between $\dot\gamma$ and $T_{\rm u}$ and between
$\dot\gamma$ and $T_{\rm sb}$ are plotted in Fig.
\ref{fig:temp}, where
we approximately found $T_{\rm u} \simeq const.$, $T_{\rm sb} \propto
\dot\gamma$.
From Eq. (\ref{eq:Tg}), we may write the relation
\begin{eqnarray}
\bar{T}_{\rm g}^\ast &=& \frac{N-n_{\rm u}}{N}T_{\rm sb}^\ast + \frac{n_{\rm
u}}{N}T_{\rm u}^\ast \nonumber \\
&=& a{\dot\gamma}^\ast - b {\dot\gamma}^{\ast2} + c, \label{temp}
\end{eqnarray}
where $\bar{T}_{\rm g}^\ast = \bar{T}_{\rm g}/\varepsilon$,
$T_{\rm sb}^\ast = T_{\rm sb}/\varepsilon$ and
$T_{\rm u}^\ast = T_{\rm u}/\varepsilon$.
Here $a, b, c$ are constants with respect to the shear rate
while they depend on the density and the dissipation rate.

\begin{figure}
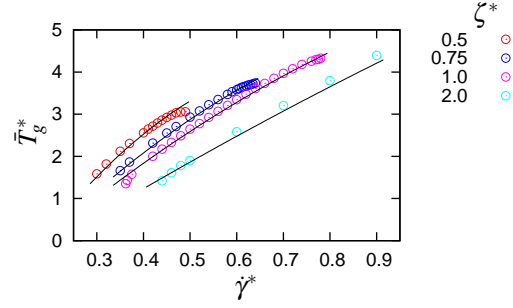

\begin{overpic}[height=.16\textheight]{Tg_of_2}
\put(80,-6){${\dot\gamma}^\ast$}
\put(-3,50){\rotatebox{90}{$\bar{T}_g^\ast$}}
\put(170,97){$\zeta^\ast$}
\end{overpic}
\caption{The dependence of the average granular temperature $T_{\rm g}$ on the shear rate $\dot\gamma$ for several dissipation rate $\zeta$ (open circles) and the results calculated by using Eq. (\ref{temp}) (solid lines).}
\label{fig:Tg}
\end{figure}
The dependence of $\bar{T}_{\rm g}$ on the shear rate $\dot\gamma$
calculated from the definition and that from Eq. (\ref{temp}) are plotted 
in Fig. \ref{fig:Tg}.
We found that the expression (\ref{temp})
well reproduces the results of our simulation.

\begin{figure}
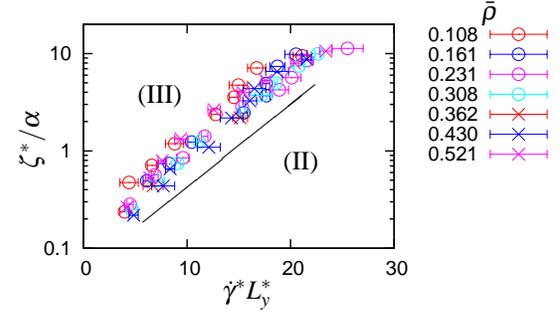

\begin{overpic}[height=.17\textheight]{phasediagram2and3}
\put(82,-4){${\dot\gamma}^\ast L_y^\ast$}
\put(5,45){\rotatebox{90}{$\zeta^\ast/\alpha$}}
\put(180,102){$\bar\rho$}
\put(105,45){(II)}
\put(50,70){(III)}
\put(52,25){\line(5,4){65}}
\end{overpic}
\caption{Boundary between (II) and (III) for each density $\bar\rho$,
where the solid line indicates $\zeta \propto \alpha \exp(\beta \dot\gamma L_y)$ with the density dependent $\alpha$.}
\label{fig:phase1}
\end{figure}

Now let us study the dependence of the phase boundary lines
among the phases on the density.
For this purpose, we investigate the behavior of
$\bar\rho=0.108$, $0.161$, $0.231$, $0.362$, $0.430$, $0.521$
($L_y=88\sigma$, $72\sigma$, $60\sigma$, $48\sigma$ $44\sigma$, $40\sigma$)
in the plane of the shear rate
$\dot\gamma$ and the dissipation rate $\zeta$ for several $\bar\rho$.
From our simulation, we found that
the phase boundary between (II) and (III) is describes by
\begin{equation}
\zeta^\ast = \alpha \exp \left( \beta {\dot\gamma}^\ast L_y^\ast \right)
\label{eq:2and3},
\end{equation}
where $L_y^\ast = L_y/\sigma$.
Here, $\alpha$ and $\beta$ are constants with respect to the shear rate and
the dissipation rate, but $\alpha$ solely depends on the density $\alpha=\alpha(\rho)$ as
\begin{eqnarray}
\alpha(\rho) &=& 453.66\rho^5 - 721.97\rho^4 + 425.66\rho^3 \nonumber\\
					&&-114.02\rho^2 +13.62\rho -0.3746.
\end{eqnarray}
Note that $\beta$ is independent of the density.
The physical reason of Eq. (\ref{eq:2and3}) can be understood as follows.
The boundary between phase (II) and phase (III) might be determined by the
condition
whether the particle trapped in the potential can pop out.
The energy gained by a shear $\delta E$ is proportional to $\dot\gamma^2
L_y^2$, and the granular temperature $T_{\rm sb}$ in the shear band is
proportional to $\dot\gamma L_y$ (the inverse temperature is proportional to $(\dot\gamma L_y)^{-1}$).
From these relations, the probability of pop out $p$ satisfies the relation
$p \propto e^{-\beta \dot\gamma L_y}$, which means $1/\tau \sim \zeta t^\ast e^{-\beta \dot\gamma L_y}$, 
where $\tau$ is characteristic time scale of pop out and $(\zeta t^\ast)^{-1}$ is that of the dissipation.
Therefore this relation becomes $\zeta \propto \exp(\beta \dot\gamma L_y)$.

\section{CONCLUSION}
\quad We have clarified the relations between the shear rate and the
dissipation rate in the system for cohesive particles
under a plane shear.
In the phase (II), we found that the dependence of the
granular temperature on the shear rate is approximately given by
$T_{\rm g} = a\dot\gamma - b\dot\gamma^2 +c$.
We also found that the boundary between (II) and (III)
is approximately described by $\zeta \propto \exp (\beta \dot\gamma L_y)$.

We suppose that the mechanism of the boundary between (I) and (II)
is related to the stability criterion for granular gas 
\cite{Saitoh2011,Alam2012,Alam1998,Alam1997,Savage1992,Garzo2006,Schmid1994,Gayen2006,Nott1999}.
This will be a future subject of our study.


\begin{theacknowledgments}
\quad Numerical computation in this work was carried out at the Yukawa Institute Computer Facility.
This work is partially supported by and the Grant-in-Aid
for the Global COE program gThe Next Generation of
Physics, Spun from Universality and Emergence from
MEXT, Japan.
\end{theacknowledgments}



\bibliographystyle{aipproc} 


\IfFileExists{\jobname.bbl}{}
{\typeout{}
\typeout{******************************************}
\typeout{** Please run "bibtex \jobname" to optain}
\typeout{** the bibliography and then re-run LaTeX}
\typeout{** twice to fix the references!}
\typeout{******************************************}
\typeout{}
}


\end{document}